\documentclass[9pt,twocolumn,twoside]{opticaStealth}
\setboolean{shortarticle}{true}
\usepackage{graphicx}   
\usepackage{verbatim}   
\usepackage{color}      
\usepackage{subfigure}  
\usepackage{hyperref}   
\usepackage{float}
\usepackage{amsmath}

\usepackage{epsfig}

\usepackage{epstopdf}
\usepackage{upgreek}
\usepackage{wasysym}
\usepackage{siunitx}
\usepackage[utf8]{inputenc} 
\usepackage{color}
\definecolor{Gray}{gray}{0.5}
\definecolor{Red}{rgb}{1.0,0.0,0.0}
\definecolor{Blue}{rgb}{0.0,0.0,1.0}
\definecolor{Green}{rgb}{0.0,1.0,0.0}

\usepackage{tikz}
\usepackage{graphicx}
\usepackage{dcolumn}
\usepackage{bm}
\usepackage{xspace}
\usepackage{gensymb}
\usepackage{soul}
\usepackage{tikz}
\usepackage{float}

\title{Coherence switching of a degenerate VECSEL for multimodality imaging}

\author[1]{Sebastian Knitter}
\author[2]{Changgeng  Liu}
\author[1]{Brandon Redding}
\author[3,5]{Mustafa K. Khokha}
\author[1,2,4,5]{Michael A. Choma}
\author[1,*]{Hui Cao}

\affil[1]{Department of Applied Physics, Yale University, New Haven, CT, USA}
\affil[2]{Department of Radiology and Biomedical imaging, Yale University, New Haven, CT, USA}
\affil[3]{Department of Genetics, Yale University, New Haven, CT, USA}
\affil[4]{Department of Biomedical Engineering, Yale University, New Haven, CT, USA}
\affil[5]{Department of Pediatrics, Yale University, New Haven, CT, USA}

\affil[*]{Corresponding author: hui.cao@yale.edu}

\dates{Compiled \today}

\ociscodes{(140.3580)  Lasers, solid-state; (110.2945) Illumination design; (170.3880) Medical and biological imaging}

\doi{\url{http://dx.doi.org/10.1364/optica.XX.XXXXXX}}

\begin{abstract}
 We demonstrate a VECSEL (vertical external cavity surface emitting laser) based degenerate source with an adjustable degree of spatial coherence that is electrically pumped, mechanically compact and supports continuous-wave emission. The laser operation can be switched between a large number of mutually incoherent spatial modes and  few-mode operation at little power loss. This technology allows multimodality imaging, where low spatial coherence illumination is used for traditional high-speed video-microscopy and high spatial coherence illumination is used to extract dynamic information of flow processes. The initial demonstration is performed on imaging embryo heart function in \textit{Xenopus}, which is an important animal model for human heart disease.
\end{abstract}

\setboolean{displaycopyright}{true}

\begin{document}

\maketitle
\thispagestyle{fancy}
\ifthenelse{\boolean{shortarticle}}{\abscontent}{}

Traditional single-mode lasers are characterized by their brightness, efficiency, and output directionality. These properties have enabled tremendous advances in imaging and sensing. However, to date, lasers have not been widely used as illumination sources for full-field imaging and display applications. This limitation exists because the high spatial coherence of existing lasers results in coherent artifacts such as speckle. In order to avoid coherent artifacts, sources with low spatial coherence typically are used. Unfortunately, traditional low spatial coherence sources such as incandescent lamps and light emitting diodes (LEDs) are not sufficiently bright for certain high-speed imaging or wide-area projection applications. One approach to overcoming the brightness limitations of traditional low spatial coherence sources is to use bright single-mode lasers in conjunction with various extracavity compounding methods to synthesize low spatial coherence light \cite{goodman2007speckle}. Although conceptually straightforward, these methods have not reached large deployment in imaging, sensing, and display. Moreover, several of these methods require multiple sequential image acquisitions or integration over a longer exposure time, thereby limiting their use in high-speed imaging. A new and promising approach for laser-based illumination focuses on intracavity design as opposed to extracavity compounding. Specifically, this design-driven approach develops laser cavities with large numbers of mutually-incoherent lasing modes ($\approx 10^3$). Consequently, the emission out of the cavity has low spatial coherence and thus mitigates or eliminates speckle artifacts. To this end, different types of laser have been developed for speckle-free imaging including dye-based random lasers \cite{redding2012speckle}, chaotic microcavity semiconductor lasers \cite{reddingPNAS15}, broad-area vertical cavity surface emitting lasers (VCSEL) \cite{riechert2008speckle,craggs2009thermally,verschaffelt2009spatially}, VCSEL arrays \cite{redding2014full}, and pulsed, solid-state degenerate (self-imaging cavity) lasers \cite{nixon2013efficient}.

In this paper we report our development of a degenerate laser with spatial coherence switching capability that is designed around a semiconductor gain element. This laser overcomes barriers presented by previous low spatial coherence lasers and demonstrates the unique potential of spatial coherence tuning for multimodal biomedical imaging. In terms of overcoming barriers, the low spatial coherence, speckle-free laser is high-power, continuous-wave (CW), and electrically-pumped. The semiconductor laser has a straightforward self-imaging design that can be built in lab without any custom fabrication. The laser is thermoelectrically cooled, so that the system can be kept compact with low maintenance cost. The gain medium is an electrically-pumped vertical external cavity surface emitting laser (VECSEL) with a large active area. 
While VECSELs are known for high output power \cite{Tropper04, Rudin:08}, prior VECSELs used a cavity design that enable single- or few-mode  operation. Our use of a degenerate laser cavity enables either a) distributing the large amount of power available to a large ($\approx 10^3$) number of mutually incoherent spatial modes or b) concentrating power to few modes by using a pinhole in the Fourier plane of the self-imaging cavity \cite{nixon2013efficient}. Without changing the pump current, the VECSEL remains stable when switching between high-coherence operation (pinhole inserted) and low-coherence operation (pinhole removed). To demonstrate the unique potential of spatial coherence switching for multimodal biomedical imaging, we use both low and high spatial coherence light generated by our VECSEL-based degenerate laser for imaging embryo heart function in Xenopus, an important animal model of heart disease. The low-coherence illumination was used for high-speed (100 frames per second) speckle-free imaging of dynamic heart structure, while the high-coherence emission was used for laser speckle contrast imaging of the blood flow \cite{fercher1981flow,boas2010lase}. 

\begin{figure}[t]
	\captionsetup{singlelinecheck=off}
	\includegraphics[width=0.9\columnwidth]{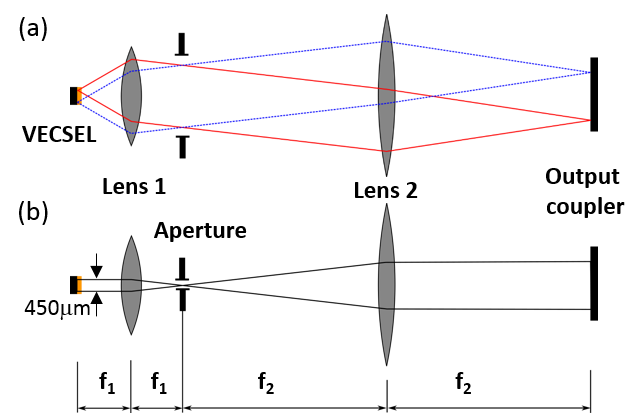}
	\caption{\label{fig:cartoon}Schematic of the degenerate laser cavity (not to scale) with: VECSEL, lens 1 (high numerical-aperture), aperture, lens 2 (reimaging lens) and output coupler (OC) mirror. In multimode operation (a) locations on the VECSEL will be imaged onto the OC and vice-versa, yielding a large number of independent spatial modes (low spatial coherence). In few-mode operation (b) an aperture in the mutual focal plane of lens 1 and lens 2, yielding plane-wave emission through the OC (high spatial coherence).}
	 \vspace{-1.5em}
\end{figure}

\begin{figure}[!b]
	\vspace{-1.0em}
		\captionsetup{singlelinecheck=off}
	\includegraphics[width=0.90\columnwidth]{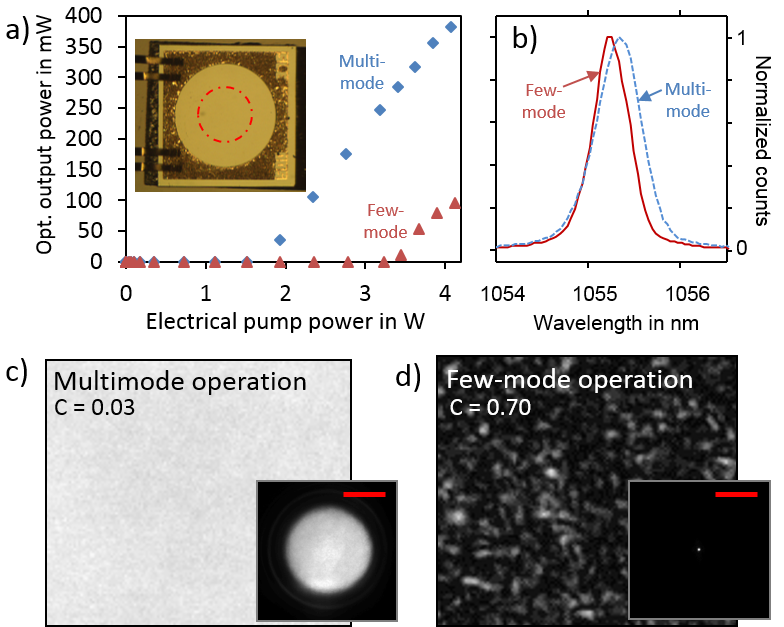}
	
	\caption{\label{fig:LaserPerf}: a) Output optical power as a function of electrical pump power of the degenerate VECSEL, with (triangles) and without (squares) a pinhole inside the cavity. Inset: Top-view of large-area VECSEL gain module with electrical contacts. Dash-dotted line highlights active region with a diameter of $450\mu m$. b) Emission spectrum of degenerate VECSEL at the electrical pump power of 4.0 W, showing a lasing peak of width $\approx0.6$ nm. c) [d)] speckle pattern produced by the output beam transmitting through a ground-glass diffuser. Low (high) speckle contrast results from the low (high) spatial coherence of laser emission. Insets: angular spectrum of the output beam recored at the back focal plane of a lens whose front focal plane coincides with the output coupler. Broad (narrow) angular spectrum indicates large (small) number of spatial modes in the multimode (few-mode) operation. Scale bars correspond to $0.4 \degree$}
\end{figure}

Figure \ref{fig:cartoon} is a schematic of the degenerate laser, which is comprised of a VECSEL gain device, two lenses, and an output coupler. The VECSEL device (made by Princeton Optronics) has an active area of diameter $\sim$ 450 $\mu$m (Fig. \ref{fig:LaserPerf} inset). It consists of multiple InGaAs/GaAs quantum wells that emit and amplify light at the wavelength $\sim$ 1055 nm with electrical pumping. The bottom distributed Bragg reflector has a reflectivity of $> 99 \%$, and serves as an end mirror for the laser cavity. The end mirror of the external cavity serves as the output coupler with a reflectivity of $95 \%$. In between the two mirrors, an aspheric singlet lens (focal length $f_1$ = 10 mm and numerical aperture NA = 0.25) and a plano-convex lens ($f_2$ = 100mm, NA = 0.1) are placed in a $2f_1$-$2f_2$ configuration. It constitutes a self-imaging system, namely, any point on one end mirror will be imaged back onto itself after light propagates one round-trip inside the cavity (see Fig. \ref{fig:cartoon}(a)). In other words, any transverse electric field distribution will be imaged onto itself after a single round trip, and therefore any field distribution is an eigenmode of the cavity. Since these eigenmodes have the same quality factor, lasing can occur simultaneously in many transverse modes. 
To reduce the number of transverse modes, a pinhole is positioned at the focal plane in between the two lenses, decreasing the resolution of the self-imaging cavity \cite{nixon2013efficient}. If the pinhole is sufficiently small, only the lowest-order transverse mode would lase (see Fig. \ref{fig:cartoon}(b)). Since the overlap of the laser mode and the gain module remains the same, the total energy, extracted from the gain module, is not changed, but redistributed among different number of transverse modes.

We demonstrate lasing with and without a pinhole in the degenerate cavity. The diameter of the pinhole is 80 $\mu$m, and it can be moved into and out of the cavity with a positioning accuracy of 1.5 $\mu$m.  Fig. \ref{fig:LaserPerf}(a) shows the output power as a function of the electrical pump power. The lasing threshold with the pinhole in the cavity is about a factor of 2 higher than that without the pinhole. This is due to the reduced overlap of the VECSEL amplifier with the sinc-shaped Fourier-image of the binary pinhole aperture. 
Figure \ref{fig:LaserPerf}(b) shows the laser emission spectra with and without the pinhole at the same electrical pump power of 4.0 W. Both are centered around 1055 nm, and have a full width at half maximum of $\sim$ 0.6 nm. The emission spectrum without the pinhole is slightly broader and red-shifted from that with the pinhole inside the cavity. As the pump power increases, the emission spectrum is shifted to longer wavelengths, due to heating effects.
To minimize the thermal effects that are common to high-power CW VECSELs, heat must be removed quickly from the VECSEL. In our setup, the gain device is mounted on a copper heat sink cooled by a thermoelectric cooler (TEC). The current of the TEC is regulated according to the temperature reading from a thermistor in close proximity of the VECSEL gain device. The heat is transferred to a finned aluminum heat sink and dissipated via forced air convection with slow-running fans.

The output beam of the degenerate VECSEL is predominantly linearly polarized with $99 \%$ and $70 \%$ of the total power in the horizontal polarization for few-mode and multimode operation, respectively. Despite the absence of polarizing elements within the laser cavity, the multiple strained InGaAs/GaAs quantum wells might provide the polarization dependent gain. 
The divergence angle of the output beam is obtained from the angular spectrum at the back focal plane of a lens, which is placed outside the cavity at a distance of one focal length from the output coupler. The half width at half maximum of the angular spectrum gives the divergence angle of $0.012 \degree$ without the pinhole [inset of Fig. \ref{fig:LaserPerf}(c)], and $0.42 \degree$ with the pinhole in the cavity [Fig. \ref{fig:LaserPerf}(d)]. After a warm-up phase of a few minutes, the laser operated very stably with relative power fluctuations of better than $0.1 \%$ (multimode) and $1 \%$ (few-mode) over the course of 1.5 hours. The larger fluctuations in few-mode operation is caused by competition between two lasing modes, which will be discussed below. 

We next demonstrate the coherence switching capabilities of our VECSEL-based degenerate laser. The spatial coherence properties of light, generated by multimode and few-mode operation, are characterized by analyzing the speckle contrast of the laser output. A polarization-maintaining optical diffuser is placed outside the cavity, and the speckle pattern of the transmitted light is detected by a CCD camera. The speckle contrast is calculated by $C = \sigma / \langle I \rangle$, where $\sigma$ is the standard deviation  and $\langle I \rangle$ is the mean of all pixel intensities within the image. As shown in Fig. \ref{fig:LaserPerf}(c), the speckle pattern without a pinhole in the degenerate laser cavity has a low contrast of $C=0.032$,  which is approximately equal to the speckle-perception threshold of the human eye \cite{wang1998speckle}. Since the laser emission bandwidth ($\sim 0.6$ nm) is much smaller than the spectral correlation width of the diffuser ($\sim 6.7$ nm), the reduction in the speckle contrast is not caused by spectral compounding.  Rather, it results from incoherent summation of many distinct speckle patterns produced by individual transverse modes that lase independently inside the degenerate cavity. From the value of $C$, we estimate the number of transverse lasing modes $N = 1/C^2 = 1030$.  This number is notably lower than the number of transverse modes supported by the (perfect) degenerate cavity, estimated from the ratio of the gain area over the diffraction limited area to be approximately $10^4$. We attribute these differences to the lens aberrations, thermal lensing effects and imperfect optical alignment of the degenerate cavity, which increase the mode size. When the pinhole is inserted to the cavity, the speckle contrast increases drastically, indicating that the laser emission becomes highly coherent (see Fig. \ref{fig:LaserPerf} (d). Since the output beam is almost linearly polarized, the contrast $C = 0.70$ indicates the emission is from two mutually incoherent transverse lasing modes.  Experimentally it is very difficult to obtain single-mode lasing in such a compact degenerate laser without significant power loss, due to thermal effects and high sensitivity of the external cavity to tiny misalignment. Nevertheless, the factor of 4 reduction in the emission power, when the number of lasing modes is switched from $\sim 1000$ to $\sim 2$, is two orders of magnitude higher than the power that could be obtained by spatial filtering of the laser emission outside the cavity.

While speckle is an artifact from the perspective that it impairs visual interpretation of an image by introducing features into an image that are not present in the imaged sample, speckle nevertheless contains information about a sample. Indeed, the contrast of local speckle patterns generated by moving scatterers contains information about the speed of those scatterers \cite{fercher1981flow}. Since coherent switching of the VECSEL-based degenerate laser enables sequential acquisition of speckled and speckle-free images of a sample, we demonstrated multimodal imaging of \textit{Xenopus} embryo (tadpole) heart physiology. Xenopus is an important animal model of human heart disease.  Speckle-free imaging yield dynamic movies of heart structure, while speckled imaging with subsequent image analysis yields dynamic maps of blood flow. Our imaging setup consisted of a conventional stereomicroscope (Olympus SZX16, NA=0.15) with a C-mounted silicon-CCD array (Allied Vision G235). Output light from the degenerate laser was coupled into a multimode fiber ($d_\textrm{core} = 600 \mu m$, length = 2.0 m). The output fiber end was imaged onto the sample plane (illumination NA = 0.2), in the center of the field of view of the microscope, to illuminate the scene in transmission. 

Speckle contrast contains information about flow speed because local speckle patterns re-randomize as new ensembles of randomly-positioned particles enter local neighborhoods of pixels. Local speckle contrast reduces in proportion to the number of distinct speckle patterns that are present during the integration period of the camera. That faster (slower) flow leads to lower (higher) speckle contrast is essentially a temporal restatement of the $C=1/\sqrt{N}$ with N being the ratio of the camera integration time over the speckle decorrelation time. In our analysis, spatial speckle contrast is measured over a 20x29 grid of square subregions (13.5x13.5 $\mu$m$^2$) across the image. The speckle images are acquired at a frame-rate of 100 frames per second (fps) with an exposure time of 9 ms/frame. To enhance the motion-induced speckle-contrast reduction, the speckle-images are averaged with their nearest-neighbors within the image sequence, leading to an effective integration time of 27 ms.

\begin{figure}[!t]
		\captionsetup{singlelinecheck=off}
	\includegraphics[width=0.90\columnwidth]{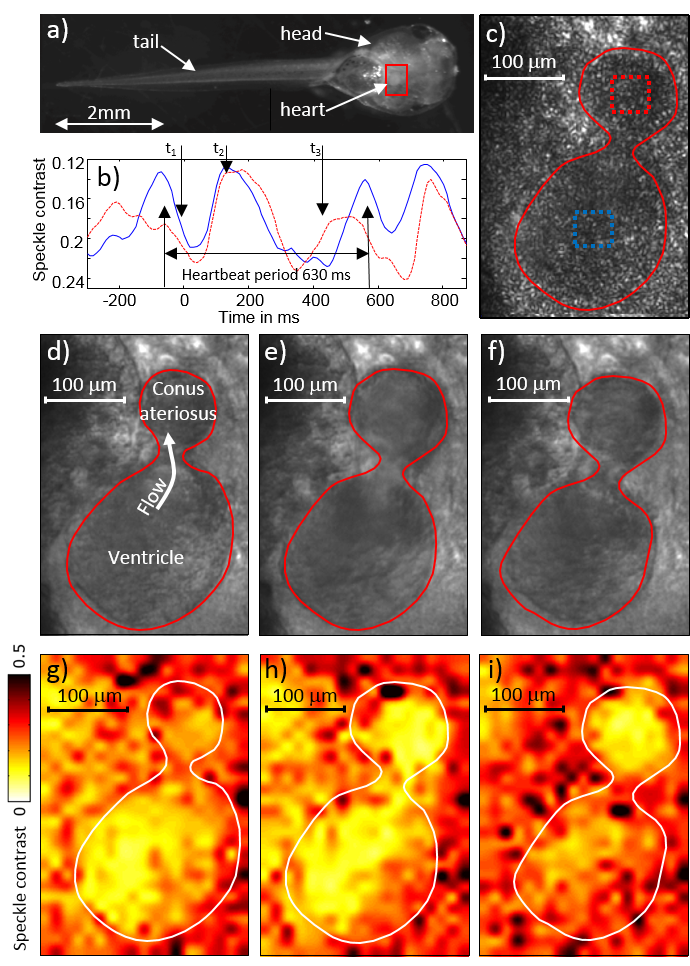}
	\caption{\label{fig:Tadpole}: a) Xenopus embryo with highlighted heart-region. b) Xenopus heart cycle, determined by speckle-contrast over time, recorded for ventricle (blue) and conus arteriosus (red).  c) Embryo heart under highly coherent illumination. (d-f) Speckle-free imaging with low-coherence illumination produces structural images of the tadpole heart with outlines of ventricle and conus ateriosus, at different phases of the heart beat. (g-i) Spatially resolved speckle-contrast, calculated from speckled images as in (c), enables the characterization of blood flow in the corresponding vessels ($20 \times 29$ segments, cubic interpolation). (d,g) ventricle is filling (diastole, $t_1$ = 0 ms); (e,h) ventricle empties into artery system (systole, $t_2$ = 150 ms), (f,i) empty ventricle, blood flows out into the vessels of the tadpole (end-systole, $t_3$ = 430 ms).}
	\vspace{-1.5em}
\end{figure}

Multimodal imaging of a Nieuwkoop and Faber stage 44 Xenopus embryo heart is shown in Fig. \ref{fig:Tadpole}. The embryo was paralyzed by Benzocaine diluted 100 times in 1/3 modified Ringers solution and at this stage of their development mostly transparent. All Xenopus procedures were reviewed and approved by Yale’s Institutional Animal Care and Use Committee, which is Association for Assessment and Accreditation of Laboratory Animal Care-accredited. Fig. \ref{fig:Tadpole}(a) shows a white-light image of the embryo under low magnification. Figure \ref{fig:Tadpole}(c) shows a speckled image of the heart generated using high spatial coherence illumination with the degenerate laser in few-mode operation. Both the heart and surrounding tissue generate strong speckle patterns. The Xenopus embryo heart has two atria, one ventricle, and a ventricular outflow that consists of the proximately located conus arteriosus and the distally located truncus arteriosus. The main heart structures visible in Fig. \ref{fig:Tadpole}(c) are the ventricle and conus arteriosus. The ventricle pumps blood into the conus arteriosus. The conus, itself contractile, receives blood from the ventricle and pumps blood into the truncus arteriosus. Blood enters the systemic circulation from the truncus. Subfigures \ref{fig:Tadpole} (d,e,f) show the tadpole heart in different stages of the cardiac cycle under low-coherence illumination. The heart starts in ventricular diastole, the part of the cardiac cycle in which the ventricle fills with blood Fig. \ref{fig:Tadpole}(d,g). During ventricular diastole, the structural (speckle-free) image shows modest increase of the outer ventricular chamber diameter (Fig. \ref{fig:Tadpole}(d)), while the local speckle contrast in the ventricle is significantly decreased (Fig. \ref{fig:Tadpole}g), indicating vigorous flow into the ventricle. Next, the heart is shown in the middle of ventricular systole, the part of the cardiac cycle in which the ventricle pumps blood into the conus arteriosus. The structural image shows a decrease in ventricle size and an increase in conus arteriosus size (Fig. \ref{fig:Tadpole}(e)). The speckle contrast image shows blood flow from the ventricle into the conus (Fig. \ref{fig:Tadpole}(h)). Last, the heart is shown at the end of ventricular systole. The diameter of the ventricle is at a minimum and the conus diameter is decreasing (Fig. \ref{fig:Tadpole}(f)). The speckle contrast image shows no flow in the ventricle and active flow in the conus (Fig. \ref{fig:Tadpole}(i)). Overall, coherence switching of the VESCEL-based degenerate laser enables high-speed biological multimodal (dynamic structural and speckle-based) flow imaging using a single laser source. Evaluating the time-dependent speckle contrast allows to trace out the heart activity in different subregions, as shown for ventricle and conus arteriosus in (Fig. \ref{fig:Tadpole}(b)).

The possibility to adjust the spatial coherence allows to record the same process in different imaging modalities. While the speckle-free image contains relevant structural information of the tadpole heart (e.g., organ boundaries), the speckle-contrast image indicates the flow activity in the different organs. In repetitive processes such as the periodically beating tadpole heart, it is possible to overlay the flow activity data with the speckle-free motion image, even though they are sequentially acquired. Since the laser is operating in continuous-wave mode, the frame-rate of such a recording is no longer limited by the repetition rate of the laser. It is therefore possible to record the beating tadpole heart at an image acquisition rate of 100 fps (9 ms exposure time). Video X in the supplementary media section shows the beating tadpole heart and the overlaid flow-activity map at a frame rate of 100 fps.

In conclusion, we present a laser system that can be toggled between highly multimodal (low spatial coherence) and few-mode (high spatial coherence) emission. Broad-area VECSEL gain-modules enable this technology and allow for electrical pumping, thermo-electric cooling and continuous-wave operation. We demonstrate the versatility of the system by performing multimodality imaging with high and low coherence illumination, being able to image the heartbeat activity of a tadpole at high frame rates. Other imaging modalities such as HiLo sectioning microscopy  \cite{santos2009optically,lim2008wide,mertz2011optical} can readily benefit from a laser-source with adjustable coherence. Such a device would alleviate complications, resulting from post-emission coherence conversion, such as necessary averaging for speckle-reduction or sacrifices in illumination intensity when enhancing the coherence. Further improvement of the performance is expected with better thermal management and more precise alignment of the external cavity. The laser coherence switching may be realized at a much higher rate by replacing the pinhole with a fast spatial light modulator, allowing high-speed alternating acquisition of speckled and speckle-free images.

We acknowledge Nir Davidson, Asher Friesem, and Micha Nixon for fruitful discussions and helpful comments. This work is supported by the National Institutes of Health (NIH) (1R21EB016163-01A1, 1R21HL125125-01A1) and the Office of Naval Research (ONR) (MURI SP0001135605).

\newpage
\onecolumn



\end{document}